\documentclass[reprint,twocolumn,groupedaddress]{revtex4-1}

\usepackage{graphicx}
\usepackage{verbatim}
\usepackage{tabulary}
\usepackage{epstopdf}

\newcommand{\cm}{cm$^{-1}$}

\begin{document}


\title{Probing the structure and dynamics of molecular clusters using rotational wavepackets}


\author{$^1$Gediminas Galinis, $^2$Cephise Cacho, $^2$Richard T. Chapman,
  $^3$Andrew M. Ellis, $^4$Marius Lewerenz, $^1$Luis G. Mendoza Luna,
  $^5$Russell S. Minns, $^4$Mirjana Mladenovi\'c, $^6$Arnaud Rouz{\'e}e,
  $^2$Emma Springate, $^2$I.~C. Edmond Turcu, $^1$Mark J. Watkins, $^1$Klaus~von~Haeften}
\email[]{kvh6@le.ac.uk} 
\affiliation{$^1$University of Leicester, Department of Physics \& Astronomy, Leicester, LE1 7RH, United Kingdom}
\affiliation{$^2$Central Laser Facility, STFC Rutherford Appleton Laboratory, United Kingdom} 
\affiliation{$^3$University of Leicester, Department of Chemistry, Leicester, LE1 7RH, United Kingdom}
\affiliation{$^4$Universit{\'e} Paris-Est, Laboratoire Mod{\'e}lisation et
  Simulation Multi Echelle, MSME UMR8208 CNRS, 5 bd Descartes, 77454 Marne-la-Vall{\'e}e, France}
\affiliation{$^5$University of Southampton, Chemistry, Southampton, SO17 1BJ, United Kingdom}
\affiliation{$^6$Max Born Institute, Max Born Strasse 2A, 12489~Berlin, Germany} 

\date{\today}

\maketitle


{\bf 
The chemical and physical properties of molecular clusters can heavily depend on
their size, which makes them very attractive for the design of new
materials with tailored properties. Deriving the structure and dynamics
of clusters is therefore of major interest in
science. Weakly bound clusters can be studied using conventional
spectroscopic techniques, but the number of lines
observed is often too small for a comprehensive structural analysis. 
Impulsive alignment generates rotational wavepackets, which provides
simultaneous information on structure and 
dynamics, as has been demonstrated successfully for {\em isolated
  molecules}~\cite{seideman1999,rosca2001,itatani2004,seideman2006,ghafur2009,bisgaard2009,ghafur2009,kanai2005,velotta2001}.     
Here, we apply this technique for the first
time to {\em clusters} comprising of
a molecule and a single helium atom. 
By forcing the population of high rotational levels in intense laser
fields we demonstrate the generation of rich rotational line spectra for
this system, 
establishing the highly delocalised structure and the coherence of
rotational wavepacket propagation.
Our findings enable studies of 
clusters of different sizes and complexity as well as incipient
superfluidity effects using wavepacket methods.} 

The study of weakly bound
clusters requires very low temperatures to allow their formation, implying
the population of only the lowest quantum levels. Consequently, 
the features in conventional rotational microwave (MW) spectroscopy
comprise only a few lines -- often not enough 
for a comprehensive analysis of the structure \cite{surin2008}.
In contrast, spectral features of variable complexity can be generated
using impulsive alignment. In this method, a rotational wavepacket forms
through the non-resonant interaction of
an intense laser field with a molecule, aligning it in
space~\cite{seideman2006}. Tuning the laser pulse duration and intensity
controls the number of rotational eigenstates in the wavepackets.
By following the evolution of alignment in time it is possible to map
these eigenstates and explore the rotational dynamics.
Given the equivalence of time and frequency-domain
information, spectra comprising of few, or many, lines can be generated. 
This ability is also of advantage for the
exploration of liquids where the wavepacket dynamics will be sensitive
to dephasing. A particularly promising liquid to begin such studies is
superfluid helium. Frequency-domain spectra of molecules 
embedded into large superfluid helium droplets show sharp 
rotational transitions, suggesting that wavepackets will not dephase
\cite{hartmann1995,surin2008}.  
However, the recent attempt to impulsively align molecules in large helium
droplets by Pentlehner~\emph {et al.} showed that this was not the
case \cite{pentlehner2013}. The findings remain unexplained.

This work was motivated to establish a bridge between free molecules and helium
droplets. We chose to study clusters of acetylene (C$_2$H$_2$) and a single
helium atom in order to reduce size and complexity to a
minimum. Rotational spectroscopy (MW) of C$_2$H$_2$ has neither been
performed in large helium
droplets nor in small clusters before, one limitation being a lack of a
permanent dipole moment needed in MW spectroscopy. Several
potential energy surfaces for C$_2$H$_2$-He nevertheless exist, allowing for comparison
between experiment and theory \cite{moszynski1995,rezaei2012,fernandez2013}.

\begin{figure}
\resizebox{1\columnwidth}{!}{\includegraphics{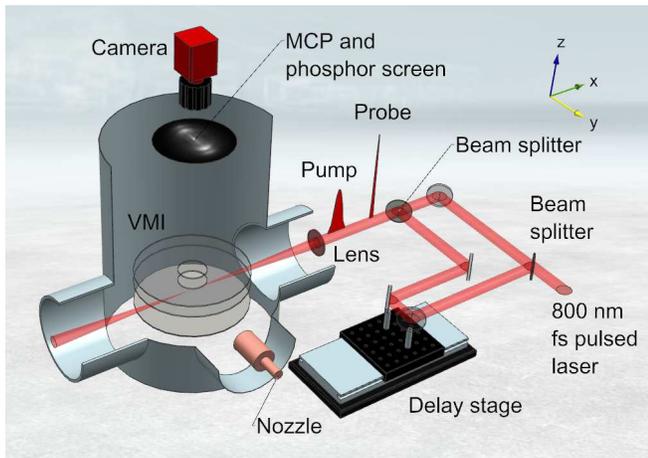}   
}
\vspace{-3mm}
\caption{
{\bf Schematic of experimental setup.} Helium-C$_2$H$_2$ clusters
produced by expansion of 0.01~\% C$_2$H$_2$ in 7~MPa He through a cooled Even-Lavie
valve were irradiated by femtosecond pulses originating from a 30~fs, 1~kHz
Ti:Sapphire laser operating at 800~nm (KM Labs Red Dragon). 
Both, pump and probe laser beams, comprising of separate grating
compressors (not shown), were colinearly focused through a $f~=~500$~mm
lens into the molecular beam 50~mm downstream from the nozzle exit.     
\label{fig:exp_setup} 
}
\end{figure}

A beam comprising small C$_2$H$_2$-He clusters was
generated in a pulsed, supersonic expansion and propagated through a vacuum
apparatus~\cite{even2000}. 
To excite rotational wavepackets 300~fs laser pulses (pump)
intersected the beam of clusters followed by 50~fs laser pulses (probe)
to detect molecular alignment (Fig.~\ref{fig:exp_setup}). 
Control of the number of rotational levels excited in the wavepackets was
achieved by variation of the intensity of the pump pulses between 5$\times$10$^{11}$ and 5$\times$10$^{12}$~Wcm$^{-2}$. The probe pulse had a constant intensity of
1$\times$10$^{15}$~Wcm$^{-2}$, sufficient to instantly break molecular bonds in a
Coulomb explosion, thereby generating C$^+$ fragment ions. Their velocity
vectors, carrying the molecular alignment information, were
selectively detected in a velocity map 
imaging (VMI) spectrometer~\cite{eppink1997} whose detector plane was
parallel to the polarisation plane of both laser beams ({\em xy} plane in 
Fig~\ref{fig:exp_setup}). The two-dimensional projection of the
recoiling C$^+$ fragment directions and intensities was used to
determine $\cos^2(\theta)$ for each position on the detector,
where $\theta$ designates the angle 
between polarisation of the pump laser and the projected velocity vectors.
The average over the entire
detector area, $\langle \cos^2(\theta)_{2D} \rangle$, is proportional to
the molecular alignment \cite{ghafur2009,pentlehner2013}. This parameter was
measured as a function of time by scanning the delay between pump and
probe laser pulses to reveal the rotational dynamics of the
clusters. 

\begin{figure}
\resizebox{1\columnwidth}{!}{%
\includegraphics[trim = 0mm 0mm 0mm 0mm, clip]{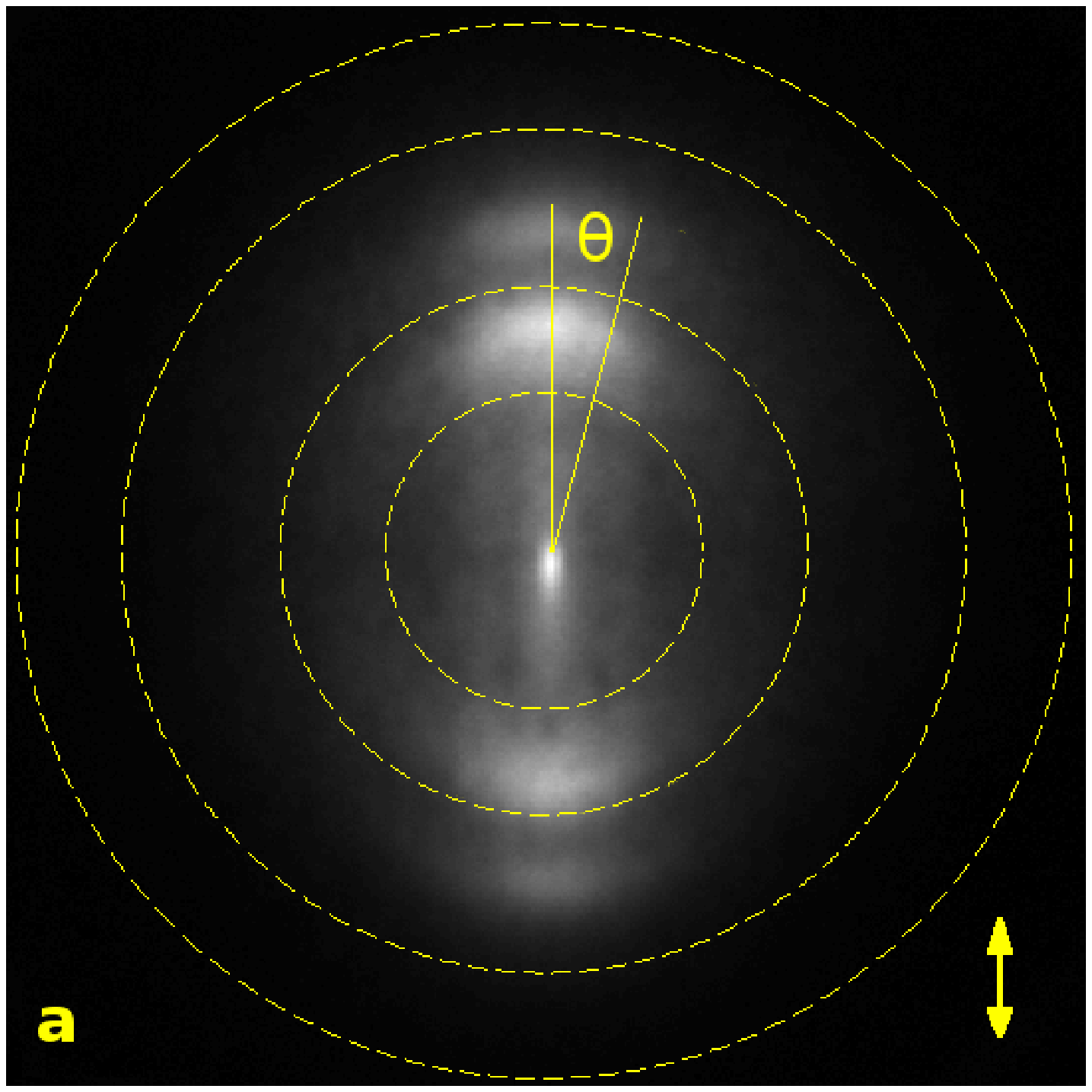}
\includegraphics[trim = 0mm 0mm 0mm 0mm, clip]{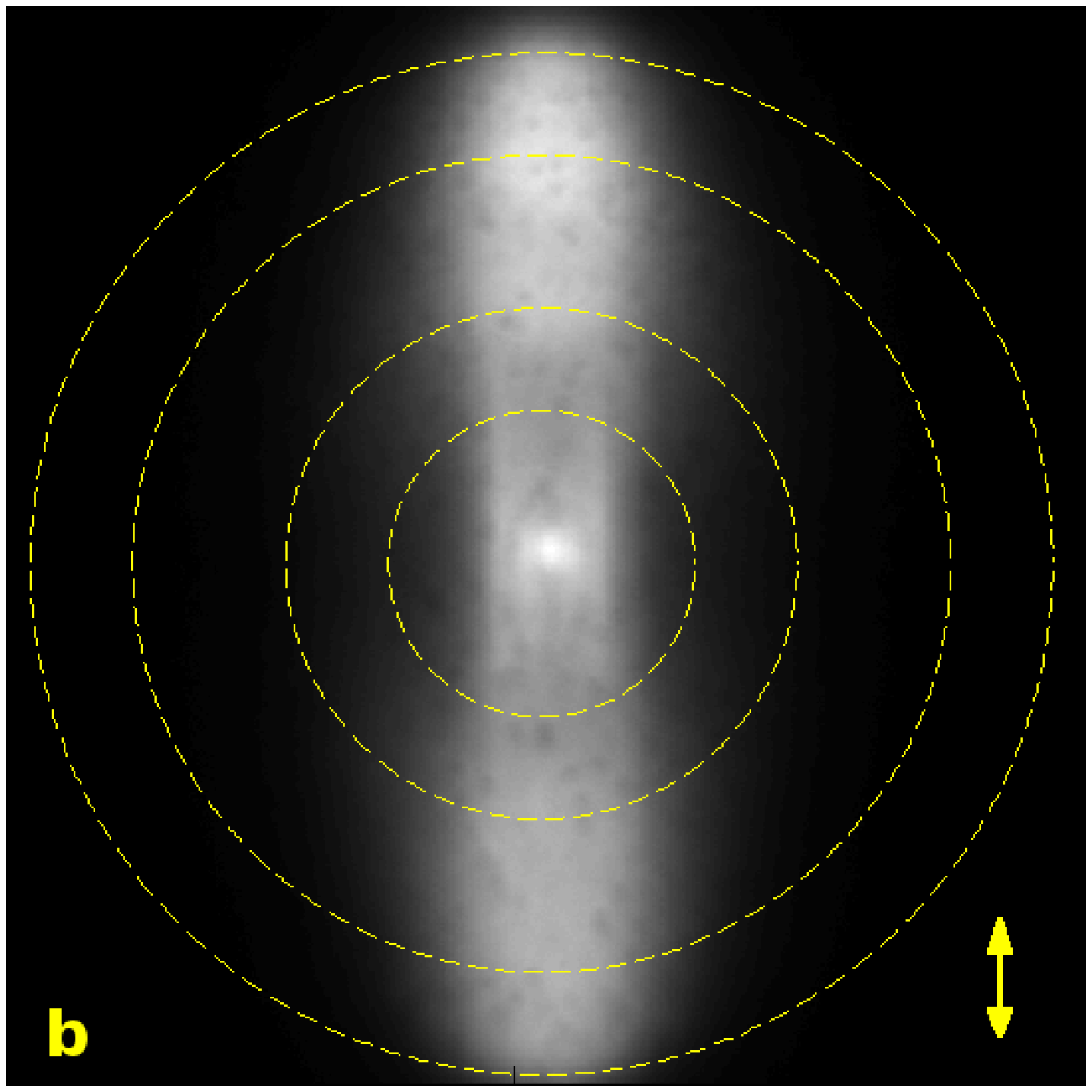}   
}
\vspace{-3mm}
\caption{{\bf Velocity map images.} Images of the Coulomb explosion process obtained via velocity
  map imaging. These particular images focus on the C$^+$ fragment and
  were recorded with different nozzle temperatures, namely {\bf (a)} 293~K and {\bf (b)}
  203~K, but under otherwise identical conditions. Production of clusters
  between helium and acetylene is
  enhanced in {\bf (b)}, supported by increased kinetic energy compared
  to {\bf (a)}. The yellow double-headed arrow 
  indicates the direction of the pump and probe laser polarisations, which are parallel to the
  {\em y}-axis in Fig.~\ref{fig:exp_setup}. The dashed circles are guides to
  the eye.
\label{fig:c_expl} 
}
\end{figure}

Fig. \ref{fig:c_expl} shows raw Coulomb explosion images 
for C$^+$ under different expansion
conditions. Fig.~\ref{fig:c_expl}(a) shows the ion image under
conditions where free C$_2$H$_2$ (no helium attached) is the dominant molecular species in the
expanding gas. This particular image shows relatively sharp features,
with evidence of more than one fragmentation channel leading to the
formation of C$^+$. By way of contrast, 
Fig.~\ref{fig:c_expl}(b) shows
the image obtained when the nozzle was cooled to 203~K to facilitate the
formation of clusters. The angular distribution of
the two images is identical, the only difference being higher kinetic
energies in image (b). Increased kinetic energy unambiguously
indicates the generation of further charges, although the laser
parameters themselves have not changed. Space-charge effects can be
excluded given that the temperature variation increases the gas number density
in the interaction region only by 30~\% -- an insignificant reduction in
the internuclear separation of ions in the 
gas compared to the high atomic number density within a cluster. We suspect that a
mechanism, similar to  
that predicted for large helium droplets doped with Xe atoms, is
operating for small clusters \cite{mikaberidze2009}. In our case the
ionisation probability of He 
atoms might be enhanced by the presence of the solvated molecule.  

\begin{figure}
\resizebox{1\columnwidth}{!}{%
\includegraphics[trim = 0mm 0mm 0mm 0mm, clip]{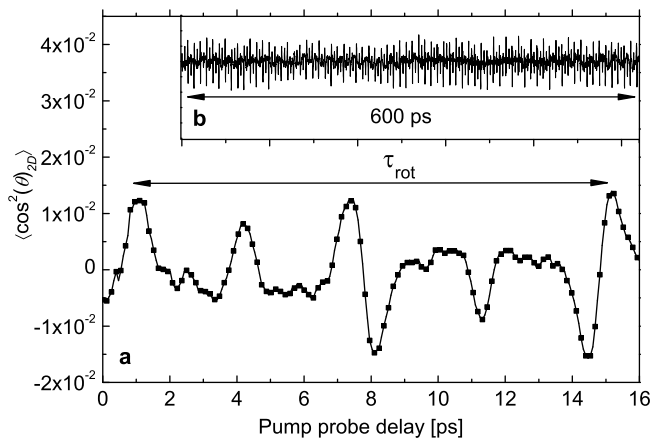}
}    
\vspace{-5mm}
\caption{
  {\bf Time-resolved alignment.} {\bf (a)} Rotational revivals obtained under
  conditions where the helium 
  stagnation pressure and temperature were 6.4~MPa and 217~K,
  respectively. Note that the baseline has been subtracted. The data has been smoothed. The time difference
  between two full revivals is equal to the rotational period of acetylene and is indicated by a double-headed arrow. {\bf (b)}
  For time delays up to the maximum of 600~ps, no decay in the amplitude,
  nor in the shape of the revivals, was observed. The 
  coherence time of the wavepacket for free C$_2$H$_2$ molecules is
  therefore at least 600~ps. 
\label{fig:pump_probe_sc} 
}
\end{figure}

The time-resolved molecular alignment, $\langle \cos^{2}(\theta)_{2D} \rangle$, is
shown in Fig.~\ref{fig:pump_probe_sc} for 
conditions similar to Fig.~\ref{fig:c_expl}(b). The strongest
contribution originates from C$_2$H$_2$ whose full, half and
quarter revivals reappear every rotational period as maximum
alignment and anti-alignment. The features extend over the
full range of the scan up to 600~ps without any appreciable decay in
amplitude or shape, so one can infer that the coherence of the
rotational wavepacket is at least 600~ps. Acetylene-helium cluster (C$_2$H$_2$-He$_n$) revivals are difficult to directly
identify in the alignment scan. Therefore, a Fourier transform is performed,
providing a much clearer picture. 
  
\begin{figure}
\resizebox{1\columnwidth}{!}{%
\includegraphics[trim = 0mm 0mm 0mm 0mm, clip]{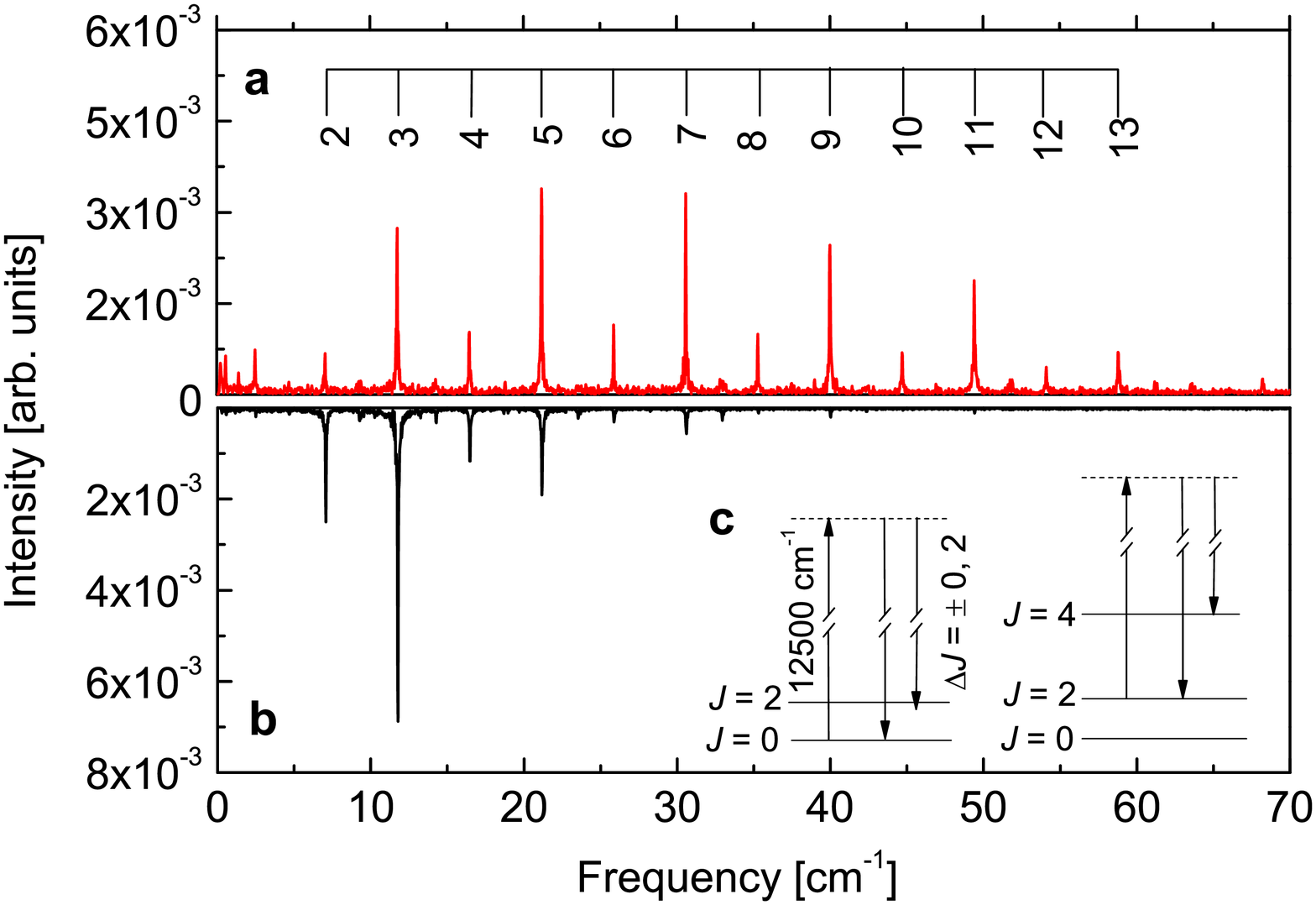}
}
\vspace{-5mm}

\caption{
  {\bf Overview spectra.} {\bf (a)}~Fourier transformation of the time-resolved
  molecular alignment recorded at a pump laser intensity of 5$\times$10$^{12}$~Wcm$^{-2}$. {\bf (b)}~Spectrum
  recorded under similar conditions, but at a lower pump laser intensity of
  5$\times$10$^{11}$~Wcm$^{-2}$. The power spectra are dominated by strong peaks
  arising from beats  
  between rotational levels of free C$_2$H$_2$ connected by
  $\Delta J=\pm 2$. The labels of the transitions correspond to the $J$
  quantum number of the final states, \emph{e.g.} '2' labels the final
  state of the $J'=2 \leftarrow J''=0$ excitation. {\bf (c)} The rotational levels
  in the wavepacket are excited through sequential Raman excitations with an
  800~nm (12,500~cm$^{-1}$) laser pulse. This process sequentially
  populates higher levels via virtual states, as schematically
  illustrated in the inset. 
  Odd $J$ levels exhibit stronger lines because of the 3:1 population
  ratio controlled by nuclear spin statistics.     
\label{fig:fft_full_range}
}
\end{figure}

The Fourier transform of the time-resolved molecular alignment is shown
for two different pump
laser intensities in Fig.~\ref{fig:fft_full_range}. 
A series of discrete lines is produced, which corresponds to particular frequency
contributions of C$_2$H$_2$ to the rotational wavepacket. The strongest features seen
in Fig.~\ref{fig:fft_full_range} coincide with the beat
frequencies of C$_2$H$_2$ at $6B$~+~$4nB$, where $B$ is the  C$_2$H$_2$
rotational constant ($B$~=~1.1769~cm$^{-1}$~\cite{callomon1957}) and
$n~=~0,~1,~2,~3,$~\emph{etc}. Six members of this series are seen at the
pump power of 5$\times$10$^{11}$~Wcm$^{-2}$. The first line at $6B$
frequency is equivalent to 
the $J'=2 \leftarrow J''=0$ transition, with successive lines
at $10B$, $14B$, $18B$, \emph{etc}. When the pump laser intensity is increased to 5$\times$10$^{12}$~Wcm$^{-2}$, higher
rotational transitions are induced, reaching up to the $J$~=~13 rotational
energy level.  

\begin{figure}
\resizebox{1\columnwidth}{!}{%
\includegraphics[trim = 0mm 0mm 0mm 0mm, clip]{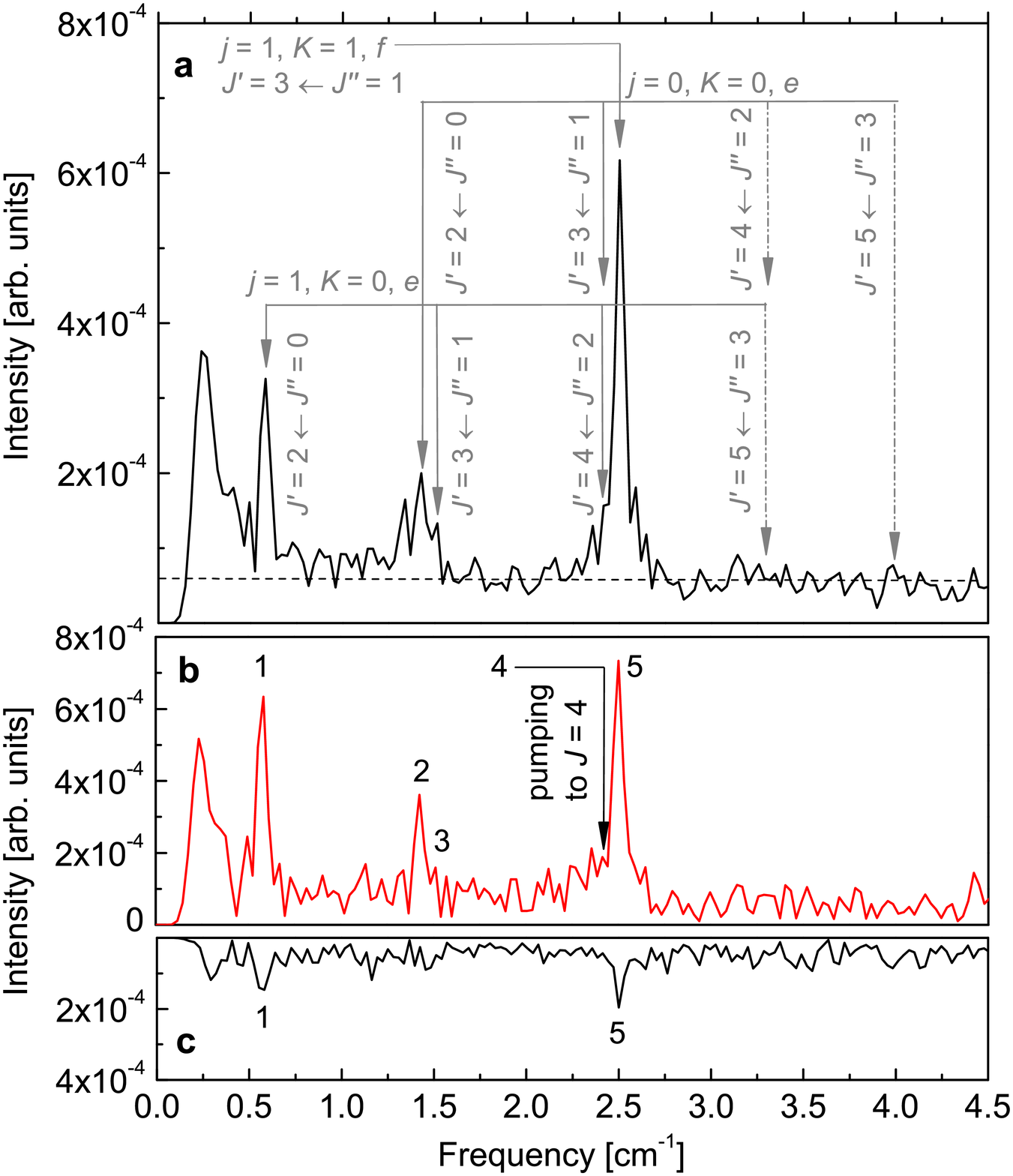}
}
\vspace{-5mm}
\caption{
  {\bf C$_2$H$_2$-He spectrum and pump laser intensity dependence.} An average of
  five spectra recorded for similar expansion conditions is shown in panel
  {\bf(a)}. Possible, but not observed transitions are marked by
  dashed-dot arrows. The noise level, established through zero pumping
  power measurements, is indicated by a horizontal dashed line. Two
  spectra at two pump laser intensities of 
  {\bf(b)}~5$\times$10$^{12}$ and {\bf
    (c)}~5$\times$10$^{11}$~Wcm$^{-2}$. Prominent C$_2$H$_2$-He features
  are annotated by numbers 1~-~5. The pump power dependence of the
  intensity of line '4' reflects sequential pumping $J'=4 \leftarrow J''=2$.  
\label{fig:fft_pump_power}}
\end{figure}

In the low frequency range, 
shown in Fig.~\ref{fig:fft_pump_power}(a), 
several prominent peaks were observed which
do not fit to the well-known rotational transitions for free
C$_2$H$_2$ and are therefore attributed to C$_2$H$_2$-He$_n$
clusters. The full widths at half maximum of these features were found to be
0.03~cm$^{-1}$, which matches the experimental limit in resolution set
by the length of the delay scan of 600~ps.  
Hence, the line
width of the peaks assigned to C$_2$H$_2$-He$_n$ clusters is consistent
with a coherence time for the rotational wavepacket of at least 600~ps.
This first observation of
coherent propagation of rotational wavepackets in small
C$_2$H$_2$-He$_n$ clusters contrasts with the strong dephasing found in large helium
droplets. As detailed below we can assign some of the C$_2$H$_2$-He$_n$ features
to the simplest cluster, C$_2$H$_2$-He.  

To guide the assignment the interaction
energies for this cluster were calculated over a wide range of
structures using the coupled cluster singles and doubles with 
perturbative triples (CCSD(T)) method. The analytical potential energy
surface was used to compute rovibrational energy levels and wave
functions, as detailed in the supplementary information. 
Selected levels and transition energies from these calculations are
displayed in Fig.~6 and collected in Table~I.
The computed rovibrational ground state energy is -7.417~cm$^{-1}$
relative to the dissociation limit into C$_2$H$_2$ + He. 
Only the total angular momentum quantum number $J$ and the parity of the
wave functions are rigorous quantum numbers.
Parity is coded by the symbols $e$ and $f$ for levels with parity
$+(-1)^J$ and $-(-1)^J$, respectively.
Inspection of the level structure and analysis of the wave functions
reveals that the C$_2$H$_2$-He 
does not behave like a linear molecule in spite of its linear
minimum energy structure. We can introduce another quantum number, $j$,
which refers to the internal rotation of the C$_2$H$_2$ unit. Although
approximate, this turns out to be a useful quantum number, which is
consistent with the floppy character of the cluster, which is completely
delocalised over the entire angular domain even in its rovibrational
ground state. We additionally label the states by the approximate quantum number $K$ for the projection of $j$ onto the
intermolecular axis.

\begin{figure}
\resizebox{1\columnwidth}{!}{%
\includegraphics{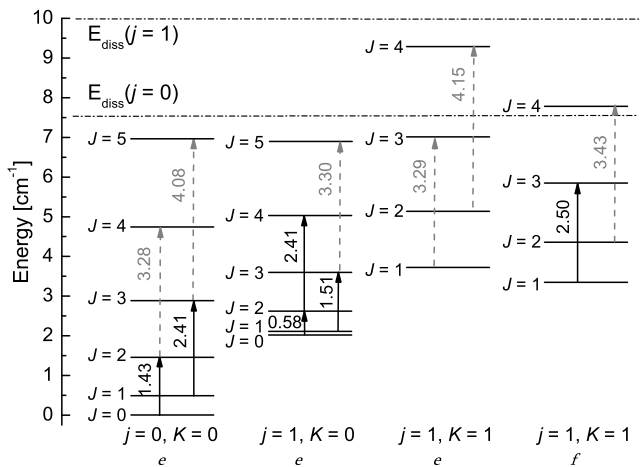}
}
\vspace{-5mm}
\caption{
{\bf Rotational energy level diagram of C$_2$H$_2$-He.} The
vertical arrows designate the transitions observed, whereas the
dashed, grey
arrows indicate possible but
not observed transitions.
\label{fig:energy_diagram} 
}
\end{figure}

\begin{table}[h!]
\begin{center}
\caption{\label{table}
Energy level assignment of measured lines. Quantum numbers
of the initial states are denoted by $j'',K'',J''$ and
of the final states by $j',K',J'$.
The experimental line positions are given in {\cm} and
the estimated error margin in each case is $\pm$0.03 {\cm}.
The transition labels in the first column are identical with the labels 
of the peaks in Fig. 5. \label{table:assignment}
}
 \vspace{0.05cm}
\begin{ruledtabular}
\begin{tabular}{crrrrrrrrrr}
 \noalign{\smallskip}
  \multicolumn{2}{c} \mbox{He-HCCH}
\\
\cline{1-2}
 \noalign{\smallskip}
Label & Sym. & $j''$ & $K''$ & $J''$ 
           &~& $j'$ & $K'$ & $J'$ 
             & Exp. & Calc. \\
 \noalign{\smallskip}
\hline
 \noalign{\smallskip}
1 & $ e$ & 1 & 0 & 0& & 1& 0& 2& 0.58 & 0.60 \\ 
3 & $ e$ & 1 & 0 & 1& & 1& 0& 3& 1.51 & 1.49 \\ 
4 & $ e$ & 1 & 0 & 2& & 1& 0& 4& 2.41 & 2.41 \\ 
 \noalign{\smallskip}
2 & $ e$ & 0 & 0 & 0& & 0& 0& 2& 1.43 & 1.46 \\ 
4 & $ e$ & 0 & 0 & 1& & 0& 0& 3& 2.41 & 2.40 \\ 
 \noalign{\smallskip}
5 & $ f$    & 1 & 1 & 1& & 1& 1& 3& 2.50 & 2.50 \\ 
 \noalign{\smallskip}
\end{tabular} 
\end{ruledtabular}
\end{center}
\end{table}

Based on these results, five peaks at 0.58, 1.43,
1.51, 2.41 and 2.50~cm$^{-1}$ were assigned
to C$_2$H$_2$-He, yielding an agreement between theory and experiment
within 0.02~cm$^{-1}$. We note that the present calculation provides a
better match to the measured line positions than the best potential
energy surface available prior to this work~\cite{fernandez2013}.  
The expected Raman transitions for  C$_2$H$_2$-He are marked in
Fig.~\ref{fig:energy_diagram} by vertical 
arrows and text, showing the 
energy differences in cm$^{-1}$. The assignment of the lines in the low
frequency range is shown in Table~\ref{table:assignment}.

The identified C$_2$H$_2$-He features depend distinctly on the pump laser intensity as shown in
Figs.~\ref{fig:fft_pump_power}(b) and (c). At lower pump intensity, two
lines exceed the noise level, namely the lines assigned to the transitions $J'=2 \leftarrow J''=0$ 
in the $j=1$, $K=0$, $e$ manifold, labelled '1', and the $J'=3 \leftarrow J''=1$
in the $j=1$, $K=1$, $f$ manifold, labelled '5'. The intensities of these lines grow with 
increasing pump laser energy due to more clusters being
excited. Additional lines emerge, for example the $J'=2 \leftarrow J''=0$ transition in the $j=0$, $K=0$, $e$ manifold, labelled '2'. 
Population transfer to higher lying states in the rotational wavepacket
due to more intense pumping is also observed. In particular, the $J''=2$
state in the $j=1$, $K=0$, $e$ manifold is pumped to the $J'=4$ state,
labelled '4'. Two further transitions, the $J'=5 \leftarrow J''=3$
in the $j=1$, $K=0$, $e$ and the $J'=5 \leftarrow J''=3$ in the
$j=0$, $K=0$, $e$ manifold, indicated in Fig.~\ref{fig:fft_pump_power}(a) by
dotted arrows, exhibit intensities within the noise level and cannot safely
be established. 

In conclusion, it has been shown for the first time that it is possible
to construct a coherent rotational wavepacket for a weakly bound
cluster between a molecule and a helium atom using impulsive
alignment. The dynamics of this wavepacket was followed in the time
domain using a second laser pulse, yielding features that can be
subjected to Fourier transformation to yield the underlying rotational
energy level structure. By this means we have established the highly delocalised, floppy structure of C$_2$H$_2$-He, a van der
Waals cluster which is only bound by \emph{ca.} 7~cm$^{-1}$. The
observed rotational energy level structure is in excellent agreement
with theoretical predictions. 

Our observation enables the simultaneous
determination of structure and dynamics in free atomic and molecular clusters, which is vital
for the fundamental understanding of size-effects and interactions in condensed
matter, for example, incipient superfluidity. This knowledge will
ultimately be critical for the development of new materials 
with tailored properties, addressing the relevance of clusters in various
applied disciplines, for example catalysis, solar cells, opto-electronics
and bio-medical imaging.


\emph{Acknowledgements}. The authors wish to thank STFC for the funded access to
the Artemis facility at the Rutherford Appleton Laboratory and the
University of Leicester for funding to support the studentship
associated with this project. Also we would like to thank S. Hook,
P. Rice, N. Rodrigues and S. Thornton for technical support during the
experiment as well as J. Underwood and M. Siano for their contributions
to the design of the VMI spectrometer. KvH kindly acknowledges funding
by STFC (seed corn fund for experiments using 4th generation light
sources) and the Leverhulme Trust (F/00212/AH). RSM would like to thank 
the Royal Society for a University Research Fellowship (UF100047), ML
and MM acknowledge computational resources funded through ANR grant
ANR-08-BLAN-0146-01.    

\emph{Author contributions}.
G. Galinis: prepared, set up and performed experiment, analysed and interpreted
data and wrote the manuscript,
C. Cacho: supported experiment,
R. Chapman: prepared, set up and supported experiment,
A. Ellis: set up and performed experiment, interpreted data and wrote manuscript,
M. Lewerenz: performed theoretical work, analysed and interpreted
data and wrote the manuscript,
L. G. Mendoza Luna: set up and performed experiment, performed theoretical work,
R. Minns: set up and performed experiment, analysed and interpreted data,
M. Mladenovic: performed theoretical work, analysed and interpreted
data,
A. Rouzee: performed theoretical work and interpreted
data, 
E. Springate: managed the laboratory and supported experiment,
E. Turcu: ran the laser system and supported experiment,
M. Watkins: set up and performed experiment, analysed and interpreted
data, performed theoretical work and wrote the manuscript,
K. von Haeften: lead the project, set up and performed experiment, interpreted
data and wrote the manuscript (principal investigator). All authors edited and commented on the manuscript.

\emph{Competing Financial Interest Statement}.
All authors declare that there are no conflicts with regards to competing
financial interests.

\end{document}


\title{
Supplementary information: \\
Prediction of rovibrational states of C$_2$H$_2$-He
}




\maketitle


The C$_2$H$_2$ unit was fixed at its experimental ground state averaged geometry \cite{herman03}.
The interaction energies between C$_2$H$_2$ and a helium atom were computed
in a Jacobi coordinate system with a Jacobi vector $R$ pointing from the
centre of mass of C$_2$H$_2$ to the helium atom and a Jacobi angle $\theta$
enclosed between $R$ and the C$_2$H$_2$ molecular axis.
Total electronic energies were computed on a grid of 300 points covering
$0 \leq \theta \leq 90^\circ$ in steps of $10^\circ$ and a
$\theta$ dependent radial grid typically ranging between 2.50 and 20~\AA.
We used the coupled cluster singles and doubles with perturbative triples, 
CCSD(T), method and included all 16 electrons in the correlation treatment.
Core optimised correlation consistent augmented (doubly augmented for He) 
basis sets according to Dunning were employed
as implemented in the MOLPRO electronic structure package 
\cite{MOLPRO_brief}.
Total energies for the complex obtained with basis sets from 
triple zeta, (d)aug-cc-pCVTZ, to quintuple zeta,
(d)aug-cc-pCV5Z, level were extrapolated to the
complete basis limit using the procedure of Peterson et al. 
\cite{peterson94}.
The extrapolated total energies were converted into interaction energies by 
subtracting the result of an extrapolation for $R=100$~{\AA} and $\theta=0$.
The grid point with the strongest interaction is at $R=4.32$~{\AA} and
$\theta=0$ with a potential value of $V_{int}=-25.100$~cm$^{-1}$
relative to separate monomers at $V=0$.
%
%
All 235 interaction energies below +200~cm$^{-1}$ were
least squares fitted to an angle dependent
extended Tang-Toennies \cite{tang84} form:
\begin{eqnarray}
   V_{int}(R,\theta) & = & A(\theta)
                       \exp\{-(b_1(\theta) R + b_2(\theta) R^2)\}
        \nonumber \\
                      &  & -\sum_{k=3,8} f_{2k}(b(\theta),R)
                                     \frac{C_{2k}(\theta)}{R^{2k}}
\end{eqnarray}
The radial parameters $X = A, b_1, b_2, C_6, C_8, C_{10}$ 
are expressed by even order Legendre series according to
\begin{equation}
      X(\theta) = \sum_{l=0,2,\ldots} X^{(l)} P_l(\theta)
\end{equation}
Higher order coefficients $C_{2k}, k>5$, are defined by
the standard recursion 
$C_{2k+2} = (C_{2k}/C_{2k-2})^3 C_{2k-4}$ and the
functions $f_{2k}$ are Tang-Toennies damping functions
with $b(\theta)=b_1(\theta)+2b_2(\theta)R$.
The fitting model contains 24 free parameters and reproduces the
ab initio interaction energies with a root mean square error of 
0.039~cm$^{-1}$ and a largest deviation of 0.22~cm$^{-1}$.
In atomic units the coefficients for this analytical representation 
have the following values:


\begin{center}
\begin{tabular}{lrp{2mm}lr}
\hline
  $A^{(0)}$     &  0.1918812E+02  && $C_6^{(0)}$    &  0.1248242E+02   \\
  $A^{(2)}$     &  0.2504029E+02  && $C_6^{(2)}$    & -0.2662071E+01   \\
  $A^{(4)}$     &  0.8700405E+01  && $C_6^{(4)}$    & -0.4200783E+01   \\
  $A^{(6)}$     &  0.1256556E+01  && $C_6^{(6)}$    & -0.5564107E+01   \\
 $b_1^{(0)}$    &  0.1327564E+01  && $C_8^{(0)}$    &  0.1318068E+04   \\
 $b_1^{(2)}$    & -0.3394156E+00  && $C_8^{(2)}$    &  0.1819554E+04   \\
 $b_1^{(4)}$    & -0.1268924E+00  && $C_8^{(4)}$    &  0.8531615E+03   \\
 $b_1^{(6)}$    &  0.1446208E-01  && $C_8^{(6)}$    &  0.1011623E+04   \\
 $b_2^{(0)}$    &  0.4538917E-01  && $C_{10}^{(0)}$ & -0.2223242E+05   \\
 $b_2^{(2)}$    &  0.4238423E-01  && $C_{10}^{(2)}$ & -0.2251989E+05   \\
 $b_2^{(4)}$    &  0.1040915E-01  && $C_{10}^{(4)}$ & -0.1157502E+05   \\
 $b_2^{(6)}$    & -0.1945577E-02  && $C_{10}^{(6)}$ & -0.3578250E+05   \\
\hline
\end{tabular}
\end{center}

Bound rovibrational levels for the fitted surface
were calculated with the DVR-DGB method,
which uses a discrete variable representation (DVR)
for the angular coordinate and a distributed
Gaussian basis (DGB) for the radial degree of freedom
\cite{mladenovic91}.
We used 51 Gauss-Legendre DVR points in $\theta$ and
an angle dependent radial basis composed of up to 85
non-evenly distributed Gaussians between 2.1 and 160~\AA.
Energy eigenvalues and wave functions were computed for both parities
and for total angular momentum $0\leq J \leq 10$, but only levels up to
$J=5$ were found to be bound.
Energy levels are converged to better than $0.001$~{\cmm}.
The computed rovibrational ground state energy is -7.417~cm$^{-1}$. 
The computed level structure and the analysis of the wave functions
shows that C$_2$H$_2$-He does not behave like a linear molecule 
in spite of its linear electronic minimum.
Already in its rovibrational ground state
the complex is completely delocalised over the entire angular domain.


%
